\documentclass[preprint]{aastex}
\usepackage[]{natbib}
\usepackage{graphics}
\usepackage{emulateapj5}
\usepackage{apjfonts}

\newcommand{\etal}{et al.}  
\newcommand{\per}{\ensuremath{^{-1}}}

\newcommand{\msun}{\ensuremath{M_{\odot}}}
\newcommand{\mbh}{\ensuremath{M_\mathrm{\bullet}}}

\newcommand{\lbul}{\ensuremath{L_\mathrm{bulge}}}
\newcommand{\kms}{km s\ensuremath{^{-1}}}
\newcommand{\sigmastar}{\ensuremath{\sigma_\star}}
\newcommand{\chisq}{\ensuremath{\chi^2}}
\newcommand{\hst}{\emph{HST}}
\newcommand{\msigma}{\ensuremath{\mbh - \sigmastar}}
\newcommand{\mgb}{\ion{Mg}{1}\emph{b}}
\newcommand{\hbeta}{H\ensuremath{\beta}}
\newcommand{\ledd}{\ensuremath{L_{\mathrm{Edd}}}}
\newcommand{\lbol}{\ensuremath{L_{\mathrm{bol}}}}
\newcommand{\hal}{H$\alpha$}

\slugcomment{}
\shorttitle{STELLAR VELOCITY DISPERSION IN MRK 501} 
\shortauthors{BARTH ET AL.}

\begin{document} 

\title{Stellar Velocity Dispersion and Black Hole Mass in
the Blazar Markarian 501}

\author{Aaron J. Barth\altaffilmark{1,2}, Luis C. Ho\altaffilmark{3},
and Wallace L. W. Sargent\altaffilmark{1}}

\altaffiltext{1}{Astronomy Department, 105-24 Caltech, Pasadena, CA
91125}
\altaffiltext{2}{Hubble Fellow}
\altaffiltext{3}{The Observatories of the Carnegie Institution of
Washington, 813 Santa Barbara Street, Pasadena, CA 91101}

\begin{abstract}
  
  The recently discovered correlation between black hole mass and
  stellar velocity dispersion provides a new method to determine the
  masses of black holes in active galaxies.  We have obtained optical
  spectra of Markarian 501, a nearby $\gamma$-ray blazar with emission
  extending to TeV energies.  The stellar velocity dispersion of the
  host galaxy, measured from the calcium triplet lines in a
  $2\arcsec\times3\farcs7$ aperture, is $372 \pm 18$ \kms.  If Mrk 501
  follows the \msigma\ correlation defined for local galaxies, then
  its central black hole has a mass of $(0.9-3.4) \times 10^9$ \msun.
  This is significantly larger than some previous estimates for the
  central mass in Mrk 501 that have been based on models for its
  nonthermal emission.  The host galaxy luminosity implies a black
  hole of $\sim6\times10^8$ \msun, but this is not in severe conflict
  with the mass derived from \sigmastar\ because the
  $\mbh-\lbul$ correlation has a large intrinsic scatter.
  Using the emission-line luminosity to estimate the bolometric
  luminosity of the central engine, we find that Mrk 501 radiates at
  an extremely sub-Eddington level of $L/\ledd \approx 10^{-4}$.
  Further applications of the \msigma\ relation to radio-loud active
  galactic nuclei may be useful for interpreting unified models and
  understanding the relationship between radio galaxies and BL Lac
  objects.

\end{abstract}

\keywords{BL Lacertae objects: individual (Mrk 501) --- galaxies:
  active --- galaxies: elliptical and lenticular --- galaxies:
  kinematics and dynamics --- galaxies: nuclei}

\section{Introduction}

The tight correlation recently discovered between stellar velocity
dispersion and black hole mass in nearby galaxy bulges \citep{fm00,
geb00a} has become the key to our understanding of black hole
demographics, as well as a new tool for probing the evolution of
galaxies and quasars.  An equally important aspect of the \msigma\
correlation is its predictive power.  While dynamical mass
measurements are observationally difficult and only possible for a
limited number of galaxies, the \msigma\ relation makes it possible to
obtain a black hole mass estimate good to $\sim40\%$ accuracy or
better from a single measurement of bulge velocity dispersion.

Having such a straightforward method to estimate black hole masses is
a tremendous boon to studies of active galactic nuclei (AGNs), because
\mbh\ is a fundamental parameter affecting the energetics and emission
properties of AGNs.  \citet{geb00b} and \citet{fer01} have recently
shown that black hole masses derived for Seyfert galaxies via the
\msigma\ relation are consistent with masses determined by
reverberation mapping, providing added confidence that AGNs do follow
the same correlation as inactive galaxies.  For some classes of AGNs,
using the correlations between \mbh\ and host galaxy properties may be
the \emph{only} reliable way to estimate the central masses.  BL Lac
objects fall in this category, since more direct methods for
determining \mbh\ (stellar dynamics or reverberation mapping) cannot
be applied.

Markarian 501 is one of the nearest known BL Lac objects.  Its
redshift is $z = 0.0337$ \citep{ulr75}, corresponding to a distance of
144 Mpc for $H_0 = 70$ km s\per\ Mpc\per.  The host of Mrk 501, also
known as UGC 10599, is a giant elliptical that appears morphologically
normal in ground-based images \citep{hic82, amc91, sfk93, wsy96,
nil99}.  Stellar absorption lines are visible in the nuclear spectrum,
along with weak emission lines (Ulrich \etal\ 1975; Stickel \etal
1993).  Mrk 501 is also a $\gamma$-ray source, and is one of the few
extragalactic objects from which TeV emission has been detected
\citep{qui96, bra97}.

In this \emph{Letter}, we present new optical spectra of Mrk 501.  The
\ion{Ca}{2} triplet lines are clearly detected, albeit substantially
diluted by the nonthermal continuum.  The stellar velocity dispersion
measured from these lines is $372 \pm 18$ \kms.  Using this result in
conjunction with the \msigma\ correlation, we derive an estimate of
the mass of the black hole in Mrk 501.

\section{Observations and Reductions}

Observations were obtained with the Palomar Hale 200-inch telescope
and Double Spectrograph \citep{og82} on 2001 June 24 UT.  On the red
side, we used a 1200 grooves mm\per\ grating blazed at 9400 \AA, and a
wavelength setting giving coverage of 8435--9075 \AA\ at 0.635 \AA\
pixel\per.  For the blue side, we used a 600 grooves mm\per\ grating
blazed at 4000 \AA, covering 4200--5950 \AA\ at 1.72 \AA\ pixel\per.
A 2\arcsec-wide slit was used for all observations.  The total
exposure time for Mrk 501 was 3.5 hours, broken into individual
30-minute exposures.

The airmass ranged between 1.01 and 1.20 during the observations, and
the slit was held at a fixed position angle of 49\arcdeg.  There were
intermittent thin clouds during the observations, and the seeing was
typically $\sim1\farcs5$.  We observed Mrk 501 again
on June 26, but due to thick clouds the data were of very poor quality
and we do not include these exposures in our analysis.  We also
observed 24 stars as velocity templates during the observing run.  The
stars included giants, subgiants, and dwarfs with spectral types
between F3 and K5.

Each frame was bias-subtracted and flat-fielded using standard
techniques.  Spectral extractions were performed using a width of
3\farcs7, corresponding to 8 pixels on the red side and 6 pixels on
the blue side.  The extractions were flux-calibrated using
observations of the standard star BD +174708, and
wavelength-calibrated using exposures of a He-Ne-Ar lamp, with a final
linear shift based on the wavelengths of night sky emission lines.
Telluric absorption bands longward of 8925 \AA\ were removed by
dividing by the normalized spectrum of the standard star, but the
corrected spectra are extremely noisy longward of 8950 \AA.  A few
narrow, weak telluric features with equivalent widths of $<0.04$ \AA\
remain in the object spectra in the region 8800-8925 \AA. These
features could not be removed by standard star division since the
standard star has intrinsic absorption features in this wavelength
range that are comparable to, or stronger than, the telluric features.

\begin{figure}
\begin{center}
\rotatebox{-90}{\scalebox{0.6}{\includegraphics{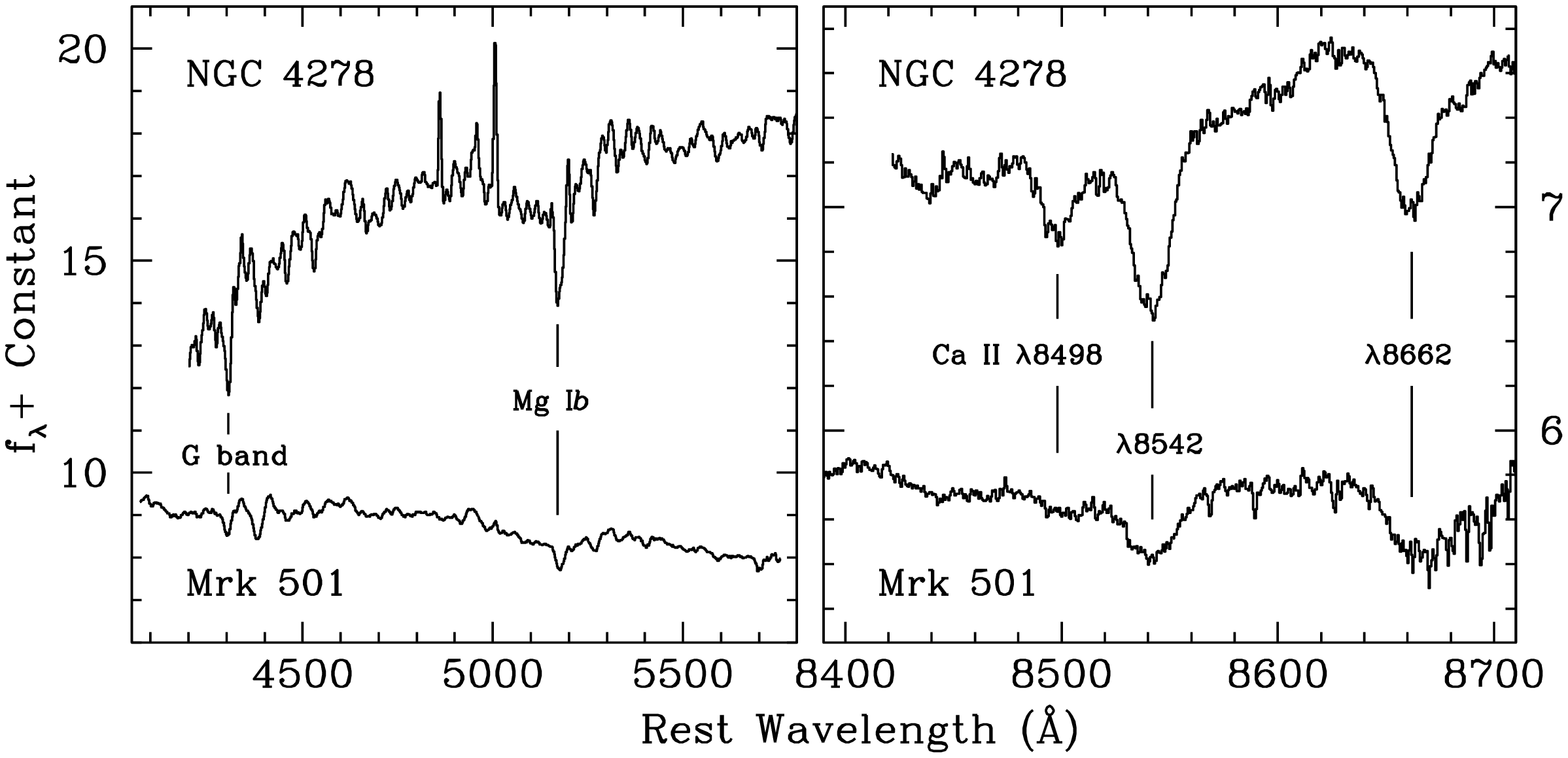}}}
\end{center}
\caption{Optical spectrum of Mrk 501.   The spectrum of the elliptical
galaxy NGC 4278, observed during the same run, is shown for
comparison.  An arbitrary constant (different for the blue and red
sides) has been added to the NGC 4278 spectrum.  
\label{figspect}
}
\end{figure}

The spectrum of Mrk 501 is displayed in Figure \ref{figspect}, along
with the nearby elliptical galaxy NGC 4278 as a comparison.  Several
stellar absorption features are clearly visible in the Mrk 501
spectrum, despite the strong dilution by nonthermal emission.  These
include the \ion{Ca}{2} triplet lines at $\lambda\lambda8498, 8542,
8662$ \AA, and on the blue side, the G band and \mgb.  Weak emission
lines of [\ion{O}{3}] $\lambda\lambda4959, 5007$ are also detected in
Mrk 501, but \hbeta\ emission is absent.  The \ion{Ca}{2}
$\lambda8542$ absorption line has an equivalent width of only $\sim1$
\AA\ (in comparison with 3--4 \AA\ in our K-giant stellar templates),
indicating that the spectrum is dominated by nonthermal emission.

\section{Measurement of Velocity Dispersion}

To measure the velocity dispersion, we used a template fitting routine
based on the method described by \citet{vdm94}, which performs a
direct fit of broadened stellar templates to the galaxy spectrum.  The
object and template spectra are first rebinned to a wavelength scale
that is linear in $\log\lambda$.  The template is broadened by
convolution with a Gaussian, and its spectral shape is adjusted with
an additive dilution, representing the nonstellar continuum, and
multiplication by a quadratic polynomial, to allow the continuum shape
of the template to match the object.  The galaxy's radial velocity is
an additional free parameter in the fit.  This procedure is repeated
using a downhill simplex search algorithm \citep{nr}, allowing all
parameters to vary freely, to find the parameter values that minimize
\chisq, yielding the best-fitting Gaussian width \sigmastar.  The
fitting routine was tested with spectra of velocity dispersion
standard galaxies from the catalog of \citet{mce95} that were observed
during the same run, and in all cases our results agreed well with the
previously measured values.

The assumption of a Gaussian for the line-of-sight velocity
distribution is a reasonable choice given the relatively low
signal-to-noise ratio of the Mrk 501 spectrum; our data do not have
the sensitivity needed to detect higher-order moments of the velocity
distribution.  The high-velocity wings on the velocity profiles due to
the central black hole should not be significant given the large
spectroscopic aperture, which corresponds to 
1.3 kpc $\times$ 2.0 kpc at the distance of Mrk 501.

\begin{figure}
\begin{center}
\rotatebox{-90}{\scalebox{0.6}{\includegraphics{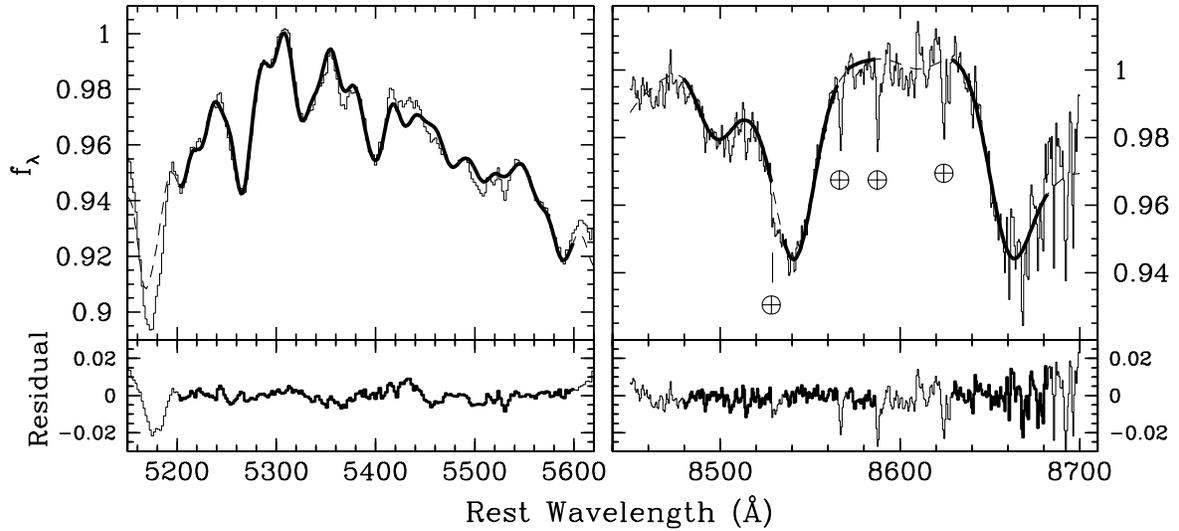}}}
\end{center}
\caption{Best fits of broadened stellar templates to the Mrk 501
spectrum. The regions included in the $\chi^2$ calculation are shown
with thick lines, and dashed lines denote the portions of the template
spectra not used in the fit.  On the red side, the strongest telluric
absorption features remaining in the spectrum are labeled.
\label{figfit}
}
\end{figure}

For the red side spectrum, we chose a fitting region just large enough
to contain the Ca lines, 8475--8680 \AA.  The spectrum becomes
extremely noisy beyond 8680 \AA\ and we truncated the fit at that
wavelength; the red wing of the $\lambda8662$ line is essentially lost
in this portion of the spectrum due to the telluric water absorption
band, even after division by the standard star.  We excluded a few
remaining weak telluric features from the fit, including one located
in the blue wing of the $\lambda8542$ absorption feature.  The
best-fitting template is HD 188056, a K3III star (Figure
\ref{figfit}).  With this template, the derived velocity dispersion is
$\sigmastar = 372 \pm 7$ \kms, and the standard deviation among all
templates observed during the run is 17 \kms.  Since template mismatch
is the dominant source of uncertainty in the measurement, we combine
the template-matching and fitting uncertainties in quadrature for a
final result of $\sigmastar=372\pm18$ \kms.

The Ca triplet region is expected to be less sensitive to stellar
population variations than the blue spectral region
\citep[e.g.,][]{dre84}, so the fit to the Ca lines should give the
best estimate of the velocity dispersion.  We also performed fits to
several portions of the blue spectrum, and found that the best results
were obtained for a region just redward of the \mgb\ line.  The \mgb\
line itself did not match the template spectra well, however, and it
was excluded from the final fits.  Over the range 5200-5600 \AA, the
best-fitting template was the K3III star HD 125560, giving $\sigmastar
= 386 \pm 9$ \kms, in reasonable agreement with the red side results.
However, several of the template stars gave very poor fits, and the
standard deviation among all templates for this region was 81 \kms.

\section{The Mass of the Black Hole in Mrk 501}

If we assume that Mrk 501 follows the \msigma\ relation defined for
nearby galaxies, then it is straightforward to convert our measurement
of \sigmastar\ into an estimate of \mbh.  This assumption is supported
by the fact that the Mrk 501 host galaxy appears morphologically
undisturbed, without any indications of recent major merger or
interaction events that might cause the galaxy to deviate from the
correlation.

The most recent fits to the \msigma\ relation are given by
\citet{kg01}, who find $\mbh \propto \sigmastar^{3.65}$, and by
\citet{mf01}, who find $\mbh \propto \sigmastar^{4.72}$.  Given the
disagreement between these two groups, the most conservative approach
is to apply both relations and derive a range of possible black hole
masses for Mrk 501.  Another issue is the fact that \cite{geb00a} use
the luminosity-weighted velocity dispersion within the half-light
radius, while \citet{fm00} base their fit on the ``central'' velocity
dispersion measured within an aperture of radius $r_e / 8$.  For our
purposes, the distinction between these two choices is not large;
\citet{geb00a} show that variations in luminosity-weighted velocity
dispersion for apertures of different size rarely exceed 10\% for $r <
4r_e$.  One possibility would be to use the prescription of
\citet{jfk95} to normalize our measured aperture dispersion to $r_e$
or $r_e/8$.  However, this would introduce additional uncertainty as
the value of $r_e$ for Mrk 501 is not well constrained by existing
ground-based imaging data.  Values for $r_e$ compiled from the
literature by \citet{nil99} range from 9\arcsec\ to 20\arcsec.  Thus,
we use our measured velocity dispersion without applying any aperture
corrections.

Propagating the uncertainty in \sigmastar\ as well as the
uncertainties in the coefficients of the \msigma\ relation, the black
hole in Mrk 501 has $\mbh = (1.3 \pm 0.4)\times10^9$ \msun\ with the
\citet{kg01} fit, or $(2.4\pm1.0)\times10^9$ \msun\ with the
\citet{mf01} fit, for a $1\sigma$ allowed range of
$(0.9-3.4)\times10^9$ \msun.  Given this result, the projected radius
of the gravitational sphere of influence ($r_g = GM/\sigma^2$) of the
black hole in Mrk 501 is expected to be $\sim0\farcs04-0\farcs16$.
This raises the possibility that a genuine stellar-dynamical mass
measurement could be done with the \emph{Hubble Space Telescope}
(\hst), if \mbh\ is in the upper half of this uncertainty range.

The main caveat to this result is that the upper end of the \msigma\
correlation is rather uncertain at present. In the range $\mbh > 10^9$
\msun, there are as yet only three galaxies with dynamical \mbh\
measurements derived from \hst\ data.  \hst\ programs currently in
progress will rectify this situation during the next few years with
many additional measurements, giving better constraints on the amount
of intrinsic scatter in the \msigma\ relation in this mass range.

\section{Discussion}

The correlation between black hole mass and galaxy bulge luminosity
can also be used to obtain black hole mass estimates.  The Mrk 501
host galaxy has a total $B$ magnitude of 14.4 (Stickel \etal\ 1993).
Correcting for Galactic extinction of $A_B=0.084$ mag \citep{sfd98}
and applying a $K$-correction of 0.16 mag \citep{pen76}, the galaxy
has $M_B = -21.6$ mag for $H_0=70$ km s\per\ Mpc\per.  From the fit to
the $\mbh - L_{\mathrm{bulge}}$ correlation given by \citet{kg01},
the expected mass is $\mbh = 6.1\times10^8$ \msun.  While this value
is outside our $1\sigma$ uncertainty range on \mbh\ from the \msigma\
relation, the scatter in the $\mbh - L_{\mathrm{bulge}}$ correlation
is more than an order of magnitude in \mbh\ at fixed
$L_{\mathrm{bulge}}$ \citep{kg01}.  Also, the host galaxy luminosity
is rather uncertain; the values compiled from the literature by
\citet{nil99} show a range of $\sim0.5$ mag among different authors.
(We note that for the adopted luminosity, the galaxy lies very close
to the mean Faber-Jackson relation for nearby ellipticals.)  Thus, the
level of disagreement between these two \mbh\ estimates is not a cause
for concern; M87 is an outlier by a similar amount in the $\mbh-\lbul$
correlation.  The \msigma\ relation is a much more reliable predictor
of \mbh\ because its intrinsic scatter is much smaller, less than
$40\%$ \citep{geb00a} and possibly near zero \citep{fm00}.

Some previous estimates of the black hole mass in Mrk 501 have been
obtained by examination of the spectrum and variability of its
nonthermal emission.  For example, \citet{fxb99} derive \mbh\ by
combining the variability timescale (from which they estimate the
physical size of the emitting region) with the assumption that the
$\gamma$-rays in blazars are produced at $\sim200$ Schwarzschild radii
from the black hole.  For Mrk 501 and seven other $\gamma$-ray
blazars, they find central masses in the range $(1-7)\times10^7$
\msun.  \citet{rm00} interpret the periodic behavior observed in X-ray
and $\gamma$-ray light curves as evidence for a binary black hole in
Mrk 501, and propose a geometric model for a jet originating from the
less massive black hole to estimate $\mbh \lesssim 10^8$ \msun\ and
$(4-42)\times10^6$ \msun\ for the two components of the binary.
Recently, \citet{wxw01} have devised a method to estimate \mbh\ in
blazars from the peak luminosity and peak frequency.  Using data for a
large sample (but not including Mrk 501), they show that their method
implies masses of $\sim10^{9-11}$ \msun\ for low-frequency peaked
blazars, and $10^{5-8}$ \msun\ for high-frequency peaked
blazars.

However, these methods have not been calibrated for galaxies whose
black hole masses have been measured dynamically, so it is difficult
to assess their accuracy.  Black hole masses below $10^8$ \msun\ in BL
Lac objects would be difficult to reconcile with the \mbh--\lbul\
correlation, since \hst\ imaging has conclusively demonstrated that
most BL Lac hosts are luminous ellipticals \citep{urr00}.  In
addition, masses of $\leq10^8$ \msun\ would pose a problem for unified
models of radio-loud AGNs, in which BL Lacs are interpreted as radio
galaxies oriented with the radio jet along our line of sight
\citep[e.g.,][]{up95}.  Nearby FR I radio galaxies, which would
presumably appear as BL Lac objects if viewed pole-on, have black
holes with masses in the range $\sim(0.3-3)\times10^9$ \msun\
\citep[e.g.,][]{har94,ffj96}.  Thus, for Mrk 501, the central mass
derived from either $\sigmastar$ or $L_{\mathrm{bulge}}$ is consistent
with the range of masses expected from unified schemes.

The Eddington ratio of the active nucleus in Mrk 501 can be determined
if its bolometric luminosity is known.  Only indirect estimates of
\lbol\ are possible, since the observed nonthermal emission is highly
beamed.  One way to estimate \lbol\ is by comparison with FR I radio
galaxies, the unbeamed counterparts of BL Lac objects.  Bolometric
luminosities for four nearby FR I radio galaxies (M84, M87, NGC 6251,
and NGC 4261) are available from \citet{ho99}, and nuclear
emission-line measurements for these objects are listed by
\citet{so81} and \citet{hfs97}.  For these four galaxies, the ratio
$L$(\hal+[\ion{N}{2}])/\lbol\ has a mean value of $4.4\times10^{-3}$
with a scatter of only $\sim25\%$ over an order of magnitude range in
\lbol.  Mrk 501 has $L$(\hal+[\ion{N}{2}]) $=1.1\times10^{41}$ erg
s\per\ (Stickel \etal\ 1993).  If its $L$(\hal+[\ion{N}{2}])/\lbol\
ratio is similar to these local FR I galaxies, as might be expected
from unified models, then it has $\lbol\approx2.4\times10^{43}$ erg
s\per\ and $L/\ledd \approx 10^{-4}$.  Mrk 501 is evidently an
extremely sub-Eddington accretor; this strengthens previous
indications that nearby TeV-emitting blazars are intrinsically weak
AGNs \citep[e.g.,][]{cm99}.

The correlations between \mbh\ and host galaxy properties make it
possible to consider new tests of AGN unification scenarios, by
examining the distributions of black hole masses and Eddington ratios
between different classes of AGNs.  Such tests have been performed
using the host galaxy luminosity to estimate \mbh\ \citep{tre01}, but
a comparison based on stellar velocity dispersion would provide
improved constraints on the central masses, and new probes of the
demographics of AGN classes such as FR I and FR II radio galaxies and
blazars.  Also, it would be worthwhile to test whether the black hole
masses differ systematically between high- and low-frequency peaked
blazars, as advocated by Wang \etal\ (2001). It should be possible to
measure stellar velocity dispersions in several other low-redshift BL
Lac objects.  At recession velocities beyond 12,000 \kms, the Ca
$\lambda8542$ line will be redshifted into telluric H$_2$O absorption
bands and difficult to measure accurately, but the \mgb\ spectral
region will still be observable.  We are beginning a spectroscopic
survey of additional nearby BL Lac objects in order to address these
questions.

\acknowledgments 

Research by A.J.B. is supported by NASA through Hubble Fellowship
grant \#HST-HF-01134.01-A awarded by the Space Telescope Science
Institute, which is operated by the Association of Universities for
Research in Astronomy, Inc., for NASA, under contract NAS 5-26555, and
by a postdoctoral fellowship (through September 2001) from the
Harvard-Smithsonian Center for Astrophysics.  Research by W.L.W.S. is
supported by NSF grant AST-9900733.  We thank Tom Matheson for writing
some of the software used for spectroscopic data reduction, Todd
Small for assistance during the observing run, and Roger Blandford for
illuminating conversations.

\clearpage

\end{document}